\documentclass[12pt,showpacs]{revtex4}
\usepackage{amssymb}
\usepackage{graphicx}
\usepackage{dcolumn}
\usepackage{bm}
\usepackage{longtable}
\usepackage{amsmath}
\usepackage{amsfonts}
\usepackage{booktabs}%
\providecommand{\U}[1]{\protect\rule{.1in}{.1in}}

\begin{document}
\title{Covariant density functional theory: The role of the pion}
\author{G. A. Lalazissis$^{1,2,3}$}%
\author{S. Karatzikos$^1$}%
\author{M. Serra$^{2}$}%
\author{T. Otsuka$^{2,4,5}$}
\author{P. Ring$^{2,3}$}%
\affiliation{$^1$Department of Theoretical Physics, Aristotle
University of Thessaloniki, GR-54124, Greece}%
\affiliation{$^2$Department of Physics, University of Tokyo, Hongo,
Bunkyo-ku, Tokyo, 113-0033, Japan}%
\affiliation{$^3$Physikdepartment, Technische Universit\"at
M\"unchen, D-85748 Garching, Germany}%
\affiliation{$^4$Center for Nuclear Study, University of Tokyo,
Hongo, Bunkyo-ku, Tokyo, 113-0033, Japan}%
\affiliation{$^5$RIKEN, Hirosawa, Wako-shi, Saitama 351-0198, Japan}


\begin{abstract}
We investigate the role of the pion in Covariant Density Functional
Theory. Starting from conventional Relativistic Mean Field (RMF)
theory with a non-linear coupling of the $\sigma$-meson and without
exchange terms we add pions with a pseudo-vector coupling to the
nucleons in relativistic Hartree-Fock approximation. In order to take
into account the change of the pion field in the nuclear medium the
effective coupling constant of the pion is treated as a free
parameter. It is found that the inclusion of the pion to this sort of
density functionals does not destroy the overall description of the
bulk properties by RMF. On the other hand, the non-central
contribution of the pion (tensor coupling) does have effects on
single particle energies and on binding energies of certain nuclei.

\end{abstract}

\maketitle

Density functional theory plays an important role for the microscopic
description of quantum mechanical many-body systems. Although this
theory is in principle exact \cite{KS.65} if the exact energy
functional is known, in nuclear physics one is far from being able to
derive this functional from underlying QCD. At the moment all
successful functionals in nuclear physics need phenomenological
parameters. It is still one of the major goals of modern nuclear
structure theory to find the optimal density functional. This will
allow reliable predictions in areas presently not yet or possibly
never accessible to experiments. This functional should be universal
in the sense that the same functional with the same set of parameters
is used all over the periodic table. In fact, by employing global
effective interactions, adjusted to properties of symmetric and
asymmetric nuclear matter, and to bulk properties of few spherical
nuclei, self-consistent mean-field models have achieved a high level
of accuracy in the description of nuclear structure properties
\cite{BHR.03}.

Relativistic versions of self-consistent mean field theories are
based on the Walecka model~\cite{SW.86}. In addition a density
dependence has been introduced~\cite{BB.77} through non-linear meson
coupling and in open shell nuclei pairing correlations have been
taken into account~\cite{VALR.05}. Two essential assumptions enter in
these descriptions: exchange-terms and vacuum polarization are not
included explicitly. This does not mean, however, that these
important contributions are neglected. Together with correlation
effects they are taken into account in a phenomenological way by the
adjustment of the parameters to experimental data.

The relativistic Hartree approximation leads to a very good
quantitative description of many nuclear properties, in particular
those of the bulk, such as binding energies, radii, deformation
parameters, giant resonances etc. The complicated structure of
effective two- and many-body forces and correlations of all sorts are
taken into account in a Lorentz invariant way by effective meson
fields, classified by their quantum numbers spin $J$, parity $P$ and
isospin $T$, as for instance the $\sigma$-meson
($J^{\pi}=0^{+},T=0$), the $\omega $-meson ($J^{\pi}=1^{-},T=0$), the
$\rho$-meson ($J^{\pi}=1^{-},T=1$), and possibly the $\delta$-meson
($J^{\pi}=0^{+},T=1$). It turns out that all these mesons have a
relatively heavy mass. In the limit of infinite mass it would be
possible to take into account the Fock terms in a Hartree theory,
using the Fierz-transformation~\cite{MBM.01}, as it is done in
non-relativistic Skyrme functionals~\cite{VB.72} since many years.
However, there is also the pion with a relatively small mass of
$m_{\pi}=138$ MeV. Certainly, it is the most important meson. It
carries the quantum numbers ($J^{\pi}=0^{-},T=1$) and is closely
connected to the chiral properties of QCD. Because of its
pseudo-scalar character it leads to a tensor force. In ab-initio
calculations based on bare nucleon-nucleon forces~\cite{WPC.00} this
part of the force gives the largest contribution to the nuclear
binding energy. A detailed study of the nucleon-nucleon
force~\cite{Mac.89} shows that indeed the $\sigma$-meson can be
understood to a large extend, by the correlated two-pion exchange. In
the nuclear medium, nonrelativistic and relativistic Brueckner
calculations~\cite{BM.90} as well as models based on chiral
perturbation theory~\cite{FKV.04} explain the largest part of the
attractive effective interaction at intermediate distances by the
contributions of the pions~\cite{HSR.07}. The pion, because of its
negative parity, does not contribute on the Hartree level. Its major
effect comes from the second and higher order diagrams in the
correlated two-pion exchange. In the phenomenological theory they are
taken into account effectively through the $\sigma$ meson. On the
tree level, however, the one-pion exchange leads to a Fock term,
which is not included in the conventional relativistic energy density
functionals based on the simple Walecka model \cite{VALR.05}. Because
of the small mass of the pion and because of the long range of the
corresponding effective force it is not clear whether a Fierz
transformation can be applied here.

In the past the pion has been included in several relativistic
Hartree-Fock calculations, as for instance in Refs:
\cite{HS.84,BMM.85,BMG.87,MNQ.91,BFG.93,MSF.04}. However, the
resulting equations of motion are rather complicated. They were
mostly solved for nuclear matter and only for a few spherical nuclei
with doubly closed shells. The numerical complexity in these computer
codes made it for a long time impossible to carry out a systematic
fit to the experimental data of many nuclei. Only recently a new
method to solve these complicated integro-differential equations in
$r$-space was implemented by Long~\textit{et al.} and improved
parameter sets have been determined~\cite{LGM.06,LSG.07} based on
density dependent meson-nucleon coupling constants.

In recent years the role of the pion in nuclei has gained renewed
interest. Although, so far, there is no direct experimental evidence,
to connect the pion with some observables, the tensor contribution of
the one-pion exchange force leads to very characteristic properties
of the single particle spectra in nuclei. It has been observed in
shell model calculations for exotic nuclei~\cite{OSF.05} that
characteristic shifts of effective single particle levels can be
traced back to the tensor interaction. It is noted that these levels
are essential to reproduce experimental data which have also been
observed directly in recent experiments by the Argonne group
\cite{SFC.04}. It has been shown~\cite{OSF.05} that the tensor force
between protons in an orbit with $j_{>}=l_{\pi}+\frac{1}{2}$ (spin
parallel to the orbital angular momentum) and neutrons in an orbit
with $j_{<}=l_{\nu}-\frac{1}{2}$ (spin anti-parallel to the orbital
angular momentum) is strongly attractive or vice versa. On the other
side it is strongly repulsive if the protons as well as the neutrons
sit in orbits with both spins parallel (or anti-parallel) to the
orbital angular momenta.

Apart from an early investigation in the seventies~\cite{SBF.77}, in
all the successful  conventional mean field calculations the tensor
force has been neglected. In most cases the parameters of those
functionals are adjusted only to bulk properties of nuclei. The
change of the single particle energies discussed above is not seen in
such calculations and therefore new versions of mean field models
have been proposed
recently~\cite{OMA.06,BDO.06,BS.07,CSF.07,LBB.07,ZDS.08,SZD.08}. Most
of these investigations are done with zero range tensor forces and in
several of them it is suggested that tensor forces have an influence
on certain single particle states around the magic numbers and that
the inclusion of the tensor force is able to improve the mean field
description and to reduce to some extent the observed deviation of
theoretical single particle levels from their experimentally measured
values~\cite{SFC.04}.

The motivation of the present work is to investigate the role of the
pion in RMF calculations and to give answers to questions such as,
(a) is the inclusion of the pion really necessary for an improved
description of data in finite nuclei, (b) what is the role of the non
central part (tensor part) of one pion exchange interaction to the
evolution of nuclear shell effects, and (c) is it necessary to
include in the density functional a tensor force for future improved
density functional theories. Having this in mind we decided to extend
the present covariant density functional and to include the pion on
the level of relativistic Hartree-Fock (RHF) theory.

\begin{table}[t]
\centering
\renewcommand{\arraystretch}{1.5}%
\caption{The parameter sets obtained after a fit with fixed value or
the pion-coupling constant $f^{2}_{\pi}= \lambda f^{2(free)}_{\pi}$,
for various values of the parameter $\lambda$.}%
\label{tab1}
\begin{center}{
\begin{tabular}
[c]{lr@{.}lr@{.}lr@{.}l}%
\hline\hline%
&\multicolumn{2}{c}{RHF(1.0)}~~ &\multicolumn{2}{c}{RHF(1.5)}~~
&\multicolumn{2}{c}{RH}\\
\hline%
$M$~(MeV)~~~~&      939&000 & 939&000 & 939&000\\
$m_{\pi}$~(MeV) &   138&000 & 138&000 & 138&000\\
$g_{\omega}$ &       13&677 &  13&588 &  12&899\\
$g_{\rho}$ &          3&606 &   4&222 &   4&589\\
\hline%
$m_{\sigma}$~(MeV)& 511&741 & 511&272 & 505&967\\
$m_{\omega}$~(MeV)& 782&501 & 782&501 & 782&238\\
$m_{\rho}$~(MeV)  & 763&000 & 763&000 & 763&000\\
$g_{\sigma}$ &       10&442 &  10&532 &  10&189\\
$g_{2}$ (fm$^{-1}$)& -7&192 &  -8&417 & -10&209\\
$g_{3}$ &           -23&248 & -26&468 & -28&612\\
\hline
$\lambda$ &           1&00  &   0&50  &   0&00\\
$\chi^{2}$ &        310&00  & 168&00  &  75&00\\%
\hline\hline
\end{tabular}
}\end{center}
\end{table}

In contrast to the investigations of Ref. \cite{LGM.06,LSG.07} we
start from the conventional form of the energy density functional
$E_{RMF}[{\hat{\rho},\phi}]$ with non-linear $\sigma$-couplings, but
no exchange terms for the mesons $\phi_m={\sigma,\omega,\rho,A}$.
However, we add the Fock term $E_{\pi}[{\hat{\rho}}]$ representing
the pseudo-vector pion propagator of Yukawa form:
\begin{equation}
E[{\hat{\rho},\phi}]=E_{RMF}[{\hat{\rho},\phi}]+E_{\pi}[{\hat{\rho}}]
\end{equation}
As discussed above, exchange terms of the heavy $\sigma$, $\omega$
and $\rho$ mesons are taken into account in this model in a
phenomenological way by adjusting the parameters of the direct terms.
Certainly this is not the case for the pion with its small mass.
Obviously, this procedure to approximate the exchange terms for the
other mesons by readjusted direct terms is a relatively economic
approximation. It facilitates greatly the numerical calculations for
the fitting procedure and, of course, enables the study of the net
effect of the pion, as everything else is left unchanged. The total
number of the parameters remains small and this to a large extent
speeds up the fitting procedure, which is obviously now considerably
more complicated due to the Fock term.

\begin{figure}[t]
\includegraphics[width=7.5cm]{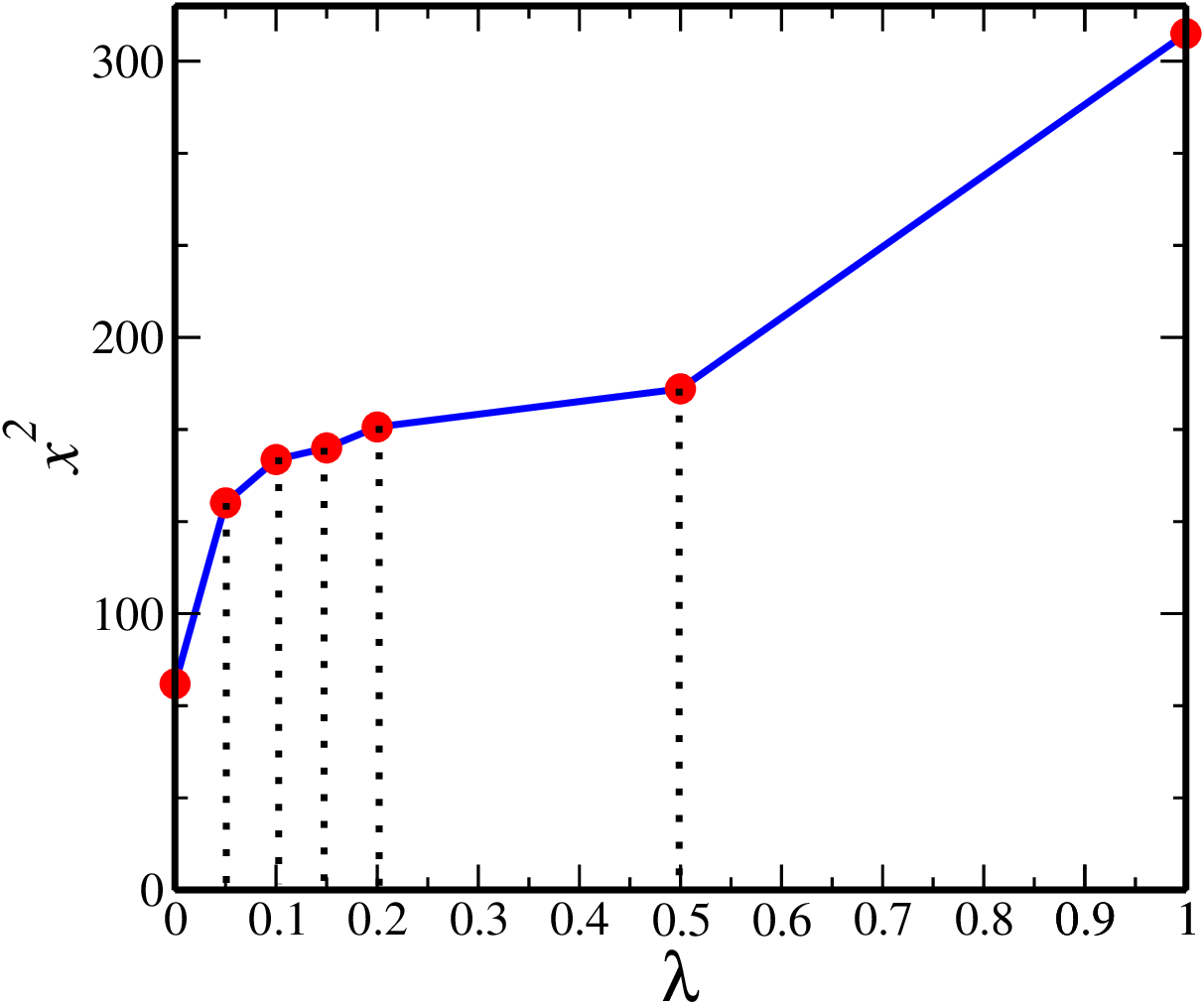}%
\caption{(Color online) $\chi^{2}$ of the fit as function of pion
coupling constant in terms of the free pion coupling constant:
$\lambda$ is defined by $f^{2}_{\pi}= \lambda
f^{2(free)}_{\pi}$}%
\label{fig1}%
\end{figure}

In this work the mass of the pion was fixed to its experimental mass
$m_{\pi}=138.0$ MeV. On the other hand the pion coupling constant, is
treated as a free parameter which varies from its experimental value
for free pions $f_{\pi}$ to zero. In other words it was considered as
an effective parameter by introducing the factor $\lambda$, with
which the $f_{\pi}$ value is multiplied.

Since the addition of the exchange terms representing the pion
changes the energy functional considerably, we have to carry out a
new fit for all the parameters entering this functional. We adopted
in the present study the philosophy and the procedure used for the
derivation of the well known parameter set NL3~\cite{LKR.97}
excluding the nucleus $^{58}$Ni. We took several fixed value for
$\lambda$ (between 0 and 1) and we performed fits for the remaining
six parameters of the model, i.e. for the mass $m_{\sigma}$ of the
$\sigma$-meson, the coupling constants $g_{\sigma}$, $g_{\omega}$,
$g_{\rho}$ and the two parameters $g_{2}$ and $g_{3}$ for the
non-linear coupling of the $\sigma$-meson. In table 1 some
representative results are shown. The RHF(1.0) set corresponds to a
fit where the full free pion coupling constant is used ($\lambda=1$).
Parameter set RHF(0.5) is the one in which tensor interaction induced
by the pion is reduced by 50\% as compared to its free value. Finally
the RH set corresponds to a fit with $\lambda=0$. In each case the
$\chi^{2}$ of the fit is calculated. The results shown in
Fig.~\ref{fig1} indicate that the model, in its present form, does
not favor the inclusion of the pion. This becomes clear from the
$\chi^{2}$ values which reflect the quality of the fit. In going from
$\lambda=1$ to $\lambda=0$, i.e as the effect of the pion is reduced,
the quality of the fit gradually improves. We investigated several
isotopic chains of spherical nuclei using the various parameter sets.
As expected, the bulk properties are not described so well in the
case of finite values of $\lambda$. Of course, one does not observe
any marked difference or peculiar behavior to these properties due to
the presence of the pion. It is interesting, however, to investigate
the influence on a microscopic level. For this purpose we studied two
cases: 1) The variation of the single particle energies of certain
levels in nuclei with magic numbers $Z$ (or $N)=20$ and $Z$ (or
$N)=28$, and 2) the variation of the energy difference between the
$1h_{11/2}$ and $1g_{9/2}$ single particle levels in Sn isotopes.

\begin{figure}[t]
\includegraphics[width=7.1cm]{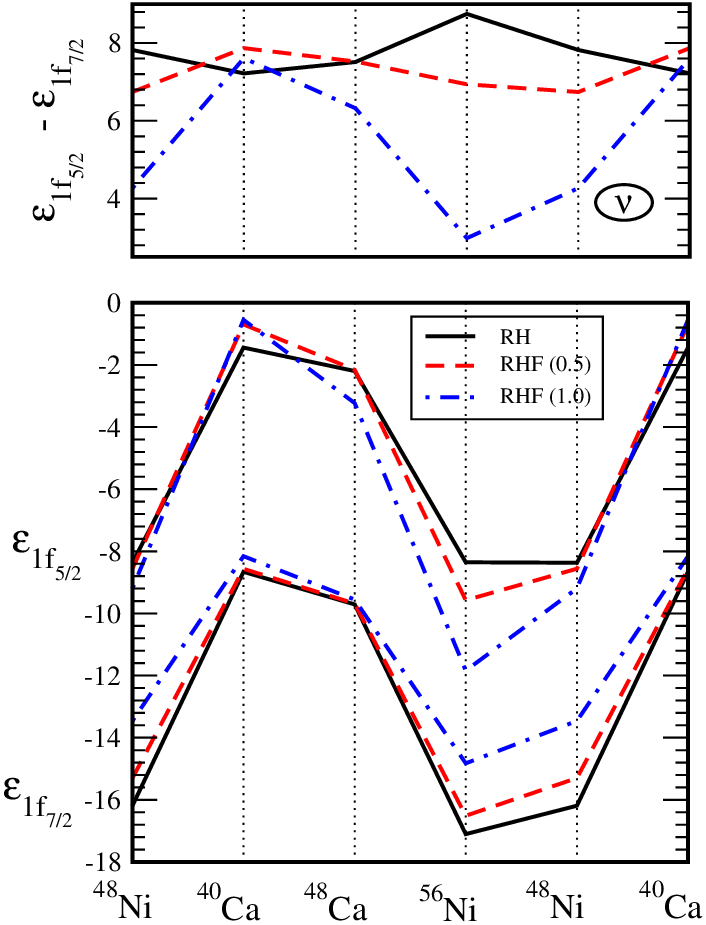}%
\caption{(Color on line) The variation of the single particle
energies (lower panel) and the spin orbit splitting (upper panel) of
the doublet $1f_{7/2}, 1f_{5/2}$ for neutrons as one moves along the
rectangular formed by the doubly magic nuclei $^{40}$Ca, $^{48}$Ca,
$^{56}$Ni, $^{48}$Ni, $^{40}$Ca.}%
\label{fig2}%
\end{figure}

\begin{figure}[t]
\includegraphics[width=7.1cm]{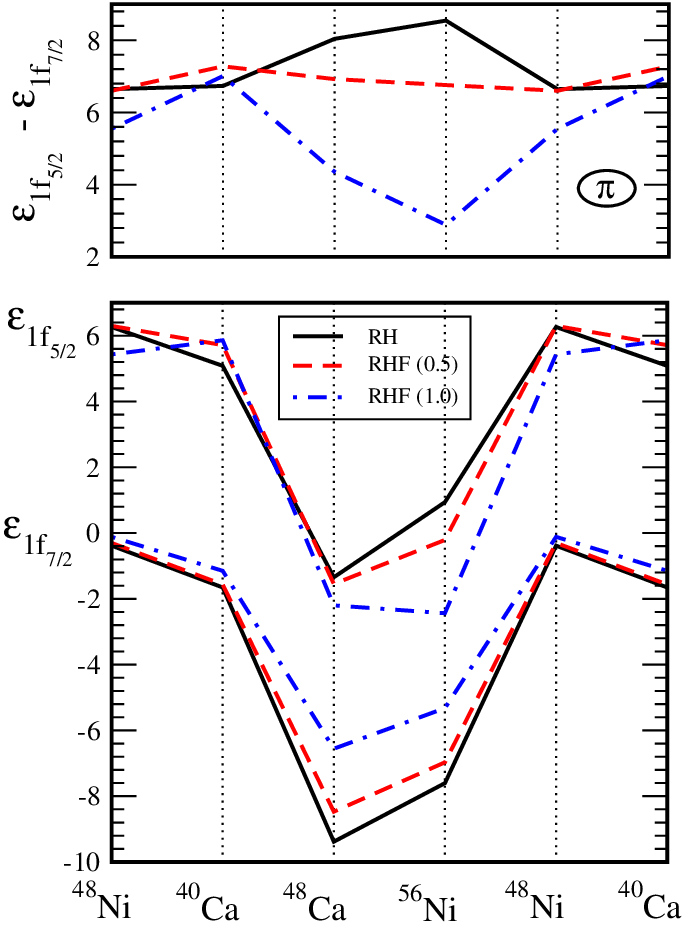}%
\caption{(Color on line) The variation of the single particle
energies (lower panel) and the spin orbit splitting (upper panel) of
the doublet $1f_{7/2}, 1f_{5/2}$ for protons as one moves along the
rectangular formed by the doubly magic nuclei
$^{40}$Ca, $^{48}$Ca, $^{56}$Ni, $^{48}$Ni, $^{40}$Ca.}%
\label{fig3}%
\end{figure}

In Figs.~\ref{fig2} and~\ref{fig3} we show the variation of the
single particle energies of the doublet $1f_{5/2},1f_{7/2}$ for
neutrons, when we go along the corners of the rectangular in the
(N,Z)-plane. We start with the doubly magic nucleus $^{40}$Ca and add
eight neutrons in the $\nu1f_{7/2}$ shell. Let us first consider
$\lambda=0$, i.e. relativistic Hartree (RH) without pions. Here we
observe relatively small shift in the single particle energies of the
neutrons, because the increasing binding caused by the balance
between $\sigma$ and $\omega$-fields is largely canceled by the
$\rho$-field, which is repulsive for the neutrons. For the protons
(Fig.~\ref{fig3}) the $\rho$-field is attractive and therefore we
find a strong lowering of the proton single particle energies when we
go from $^{40}$Ca to $^{48}$Ca. Taking into account the Fock term
caused by the pion tensor with a weight of 50 \% (dashed lines in
Figs.~\ref{fig2} and \ref{fig3}) or with a weight of 100 \%
(dashed-dotted lines) we find nearly no change for the neutron single
particle levels, but a dramatic change for the proton single particle
levels. The $\pi1f_{7/2}$ is shifted upwards as compared to the pure
Hartree calculation (full line), because, as we have seen the
interaction between the neutrons in the $\nu1f_{7/2}$ shell and the
proton in the $\pi1f_{7/2}$ shell is repulsive. On the other side the
$\pi1f_{5/2}$ orbit is shifted downward with respect to the pure
Hartree calculation (full line), because the interaction between the
neutrons in $\nu1f_{7/2}$ shell and the proton in the $\pi1f_{5/2}$
orbit is attractive. The changes follow the trends suggested in
Ref.~\cite{OSF.05}.

\begin{figure}[t]
\includegraphics[width=7.5cm]{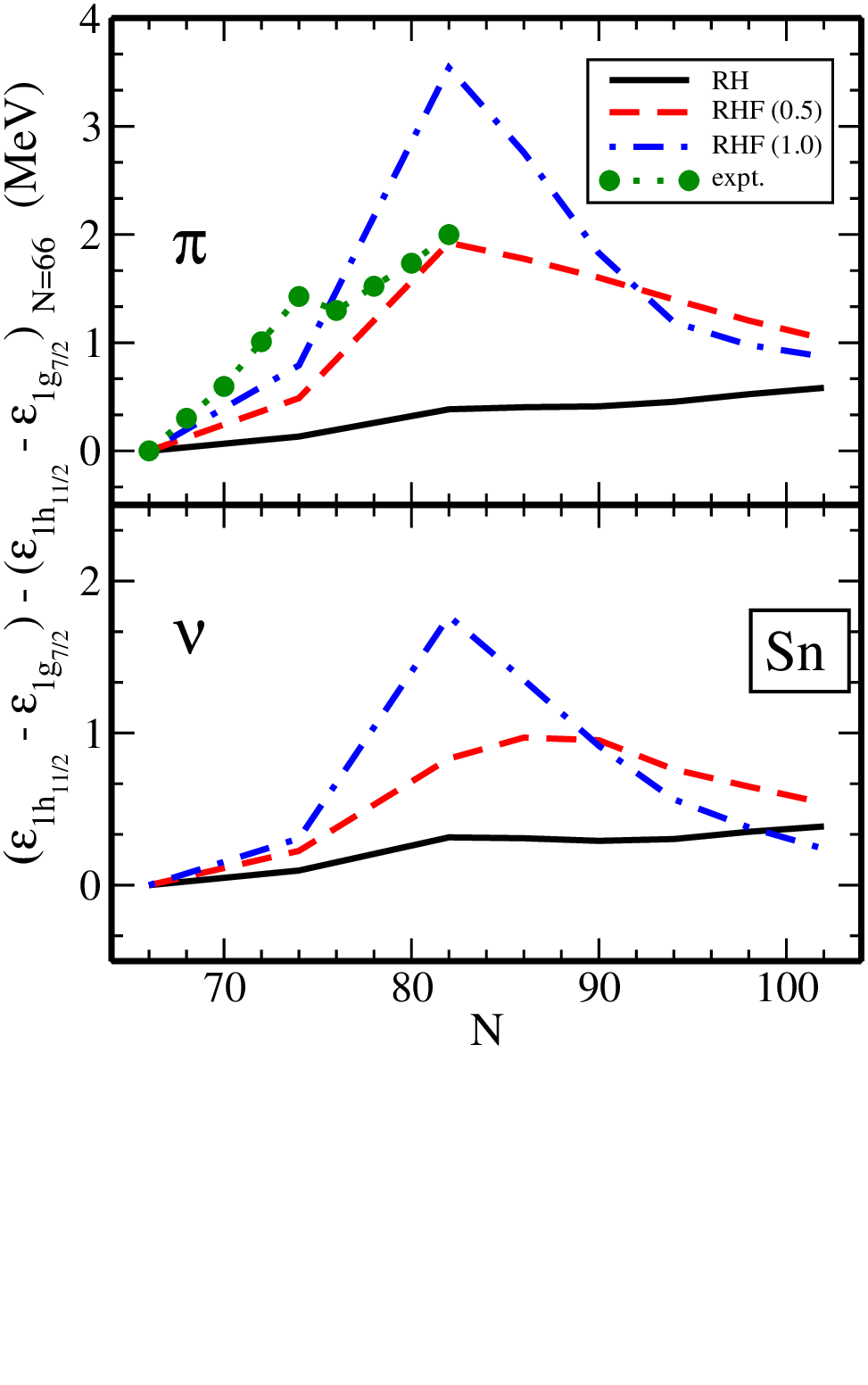}%
\vspace{-3.0cm}%
\caption{(Color online) The evolution of the
$1h_{11/2}$ and $2g_{9/2}$ neutron and proton energy gap.
The s.p levels of $^{116}$Sn are taken as reference point.}%
\label{fig4}%
\end{figure}

Next we start with the doubly magic nucleus $^{48}$Ca and add eight
protons in the $\pi1f_{7/2}$ shell until we reach the nucleus
$^{56}$Ni. In this case we observe an analogous behavior. In
relativistic Hartree (RH) without pions the proton levels change only
slightly, because of the compensating effects of the $\rho$-field.
However, the neutron levels are considerably shifted downwards.
Including the tensor field of the pions we observe again a repulsion
for neutrons in the $\nu1f_{7/2}$ shell and an attraction for the
neutrons in the $\nu1f_{5/2}$ shell, because in the first case both
protons and neutrons have the configuration $j_{>}$, whereas in the
second case the neutron $j_{<}$ and the proton has $j_{>}$, as it was
predicted in Ref.~\cite{OSF.05}.

Recently the energy difference between $\pi 1h_{11/2}$ and the $\pi
1g_{7/2}$ proton orbits in Sb-isotopes has been measured by the
Sn($\alpha,t$) reaction~\cite{SFC.04} as a function of the neutron
excess and it has been found that this difference increases steadily
with the filling of the $\nu 1h_{11/2}$-orbit. This has been
attributed to the tensor interaction in Refs.~\cite{OSF.05,BDO.06},
because this interaction is repulsive between the neutrons in the
$\nu 1h_{11/2}$-orbit ($j_>$) and the protons in the $\pi
1h_{11/2}$-orbit ($j_>$) whereas it is attractive between the
neutrons in the $\nu 1h_{11/2}$-orbit ($j_>$) and the protons in the
$\pi 1g_{7/2}$-orbit ($j_<$). In Fig.~\ref{fig4} we show the
variation of this energy difference $E_{1h_{11/2}}$ $-$
$E_{1g_{9/2}}$ on top of Sn isotopes as a function of the neutron
number. The upper panel corresponds to the single particle levels of
protons while the lower to those of neutrons. The s.p energies of Sn
isotope with N=66 are taken as a reference point. In the case where
the pion field is not taken into account (RH) this energy difference
remains practically constant as the neutron number increases. The
inclusion of the pion, however, changes the picture. As more neutrons
are added the energy gap increases and from $^{116}$Sn to $^{132}$Sn
the increase is of the order of 2-3 MeV. The effect grows steadily
with the strength of the pion coupling and therefore it may be
attributed to the tensor coupling in the pion nucleon interaction.
The results for $\lambda=0.5$ (RHF(0.5)) are in agreement with the
experimental findings of Ref.~\cite{SFC.04}. As we see in the lower
panel of Fig.~\ref{fig4} the effect seems to be less pronounced in
the case of neutrons, as expected from the isospin dependence of the
monopole part of the tensor force~\cite{OSF.05}.

\begin{figure}[t]
\includegraphics[width=7.2cm]{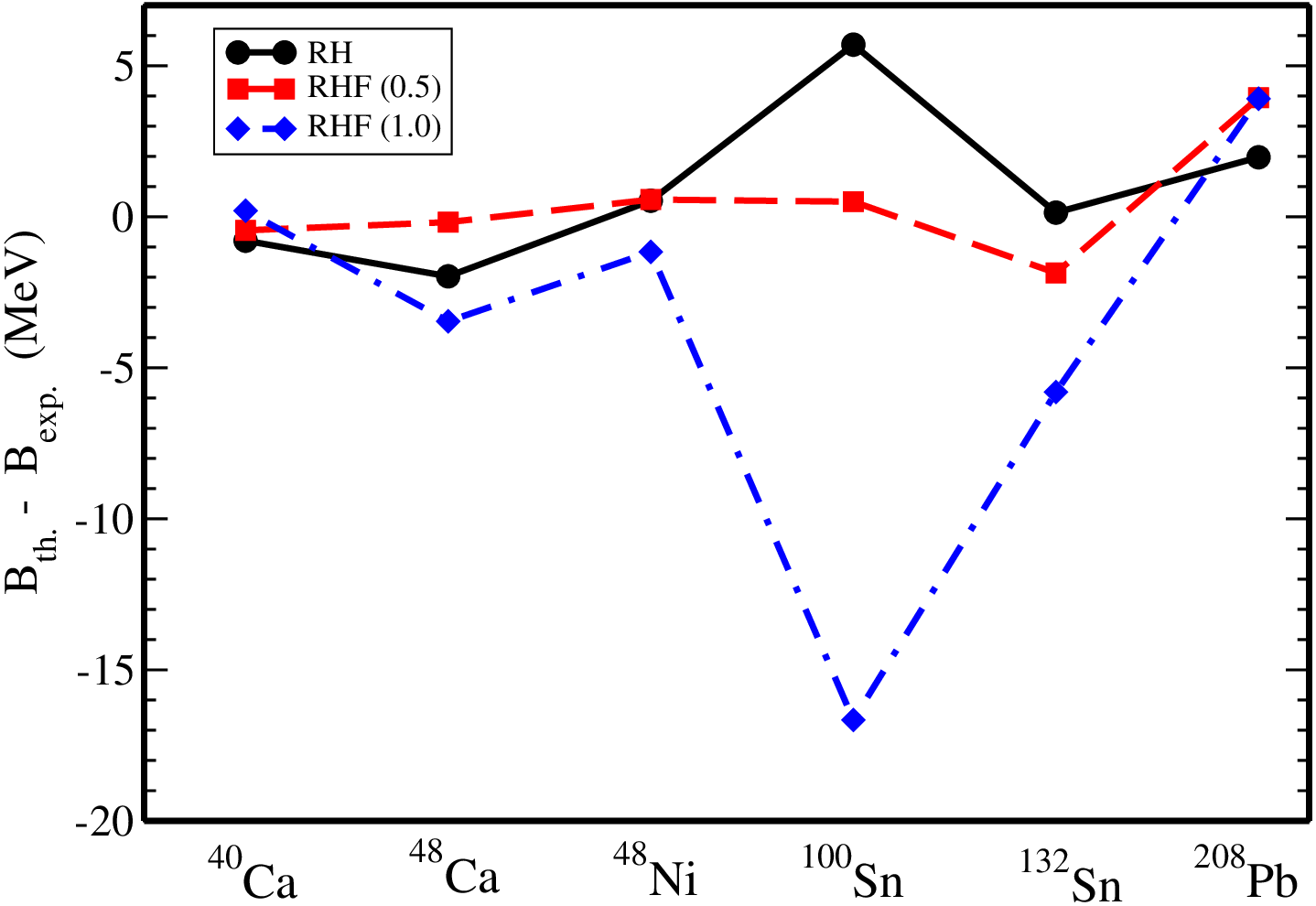}%
\caption{(Color online) Differences between theoretical and
experimental binding energies in doubly magic nuclei. RH calculations
without the pion are compared with RHF calculations including the
pion with the strength parameters $\lambda=0.5$ and
$\lambda=1.0$.}%
\label{fig5}%
\end{figure}

In Fig.~\ref{fig5} we study the influence of the tensor interaction
on the binding energies $B$ of doubly closed shell nuclei. We show
the deviations $B_{th}-B_{exp}$ of theoretical results from
experiment for various values of the strength parameter $\lambda$.
Here we have to take into account that the binding energies of the
nuclei $^{40}$Ca, $^{48}$Ca, $^{132}$Sn, and $^{208}$Pb have been
used in the fit (see Table I of Ref.~\cite{LKR.97}). Therefore the
deviations of the theoretical results from experimental data are all
relatively small for these nuclei and we find big deviations only for
the nucleus $^{100}$Sn. The RMF-results without the pion produce an
over-binding of $5.7$ MeV. The tensor force plays in the nucleus
$^{100}$Sn an important rule, because the $1g_{9/2}$-orbits ($j_>$)
for protons and neutrons are both occupied by ten particles and the
corresponding spin orbit partners $1g_{7/2}$-orbits ($j_<$) are
empty. This force is repulsive and therefore we observe with
increasing tensor force a reduced binding. For $\lambda=0.5$ we are
close to the experimental value.

In this work we have extended Covariant Density Functional Theory to
include the pion degree of freedom on the Hartree-Fock level. This
leads to tensor forces of finite range. In contrast to earlier
investigations in Refs.~\cite{LGM.06,LSG.07}, where in the nuclear
interior an exponentially decreasing pion-nucleon coupling was used,
we have adjusted the parameters of the remaining non-linear
RMF-Lagrangian for various coupling strengths of the pion field by
extensive multi parameter $\chi^{2}$ minimization procedures to
reproduce bulk properties of infinite nuclear matter and spherical
finite nuclei. The optimal fit is achieved for vanishing pion field.
However, when considering the single particle levels in semi-magic
nuclei, we observe, for finite pion fields, shifts in the particle
levels in a consistent manner with the corresponding shell model
calculations. It turned out that we can reproduce for roughly half
the strength of the tensor force produced by the free pion the
increasing energy difference between the $\pi 1h_{11/2}$ and the $\pi
1g_{7/2}$ orbits observed in recent Sn($\alpha,t$)-reactions in
Argonne~\cite{SFC.04}. We also observe an influence of the tensor
force on the binding energy in closed shell nuclei, where only one of
the spin-orbits partners is occupied. Of course, so far, we stay on
the mean field level, i.e. we do not consider coupling to low-lying
surface phonons leading to an energy dependent self energy and a
fragmentation of the corresponding single particle
energies~\cite{LR.06}. Work in this direction is in progress.

\bigskip

{\leftline{\bf ACKNOWLEDGEMENTS} This work was carried out as a
JSPS-DFG joint project. One of the authors (M.S.) acknowledges the
postdoctoral fellowship from JSPS. This work has been supported by
the DFG cluster of excellence \textquotedblleft Origin and Structure
of the Universe\textquotedblright\ (www.universe-cluster.de), by the
JSPS Core-to-Core project, EFES, and by the Hellenic State
Scholarship foundation (IKY).}


\end{document}